\documentclass[12pt]{spieman}  
\usepackage{amsmath,amsfonts,amssymb}
\usepackage{graphicx}
\usepackage{setspace}
\usepackage{tocloft}
\usepackage{subfig}
\usepackage{svg}
\usepackage{cleveref}
\usepackage{tabularx}

\title{The Effectiveness of a Simple Helmholtz coil-like Magnetic Shield at Reducing X-ray-like Background in Space-based X-ray Detectors}

\author[a,b,*]{Dr. Christopher S. W. Davis}
\author[b]{Dr David Hall}

\affil[a]{University of Surrey, Space Environment and Protection Group, Surrey Space Center, Stag Hill, University Campus, Guildford, United Kingdom, GU2 7XH}

\affil[b]{The Open University, The Center for Electronic Imaging, Robert Hooke Building, Kents Hill, Milton Keynes, United Kingdom, MK7 6AA}

\cftpagenumbersoff{figure}
\cftpagenumbersoff{table} 
\begin{document}
\maketitle

\begin{abstract}
Both active and passive magnetic shielding have been used extensively during past and current X-ray astronomy missions to shield detectors from soft protons and electrons entering through telescope optics. However, simulations performed throughout the past decade have discovered that a significant proportion of X-ray-like background originates from secondary electrons produced in spacecraft shielding surrounding X-ray detectors, which hit detectors isotropically from all directions. Here, the results from Geant4 simulations of a simple Helmholtz coil-like magnetic field surrounding a detector are presented, and it is found that a Helmholtz coil-like magnetic field is extremely effective at preventing secondary electrons from reaching the detector. This magnetic shielding method could remove almost all background associated with both backscattering electrons and fully absorbed soft electrons, which together are expected to account for approximately two thirds of the expected off-axis background in silicon-based X-ray detectors of several hundred microns in thickness. The magnetic field structure necessary for doing this could easily be produced using a set of solenoids or neodymium magnets providing that power requirements can be sufficiently optimised or neodymium fluorescence lines can be sufficiently attenuated, respectively.

\end{abstract}

\keywords{X-ray, background, astronomy, magnetic, shielding}

{\noindent \footnotesize *Dr. Christopher S. W. Davis\footnote{The work described in this article was previously presented in the thesis of the first author in Chapter 10 of Davis (2021)\cite{davis2021simulation} , and has been adapted for this article. All of the research and writing presented in this article was performed in the spare time of the first author or prior to their employment at the University of Surrey.}, \linkable{c.s.davis@surrey.ac.uk} }

\begin{spacing}{2}   

\section{Introduction}
It has been known for at least a decade that secondary electrons generated by incident protons passing through surfaces surrounding a space-based X-ray detector can contribute a significant proportion of X-ray-like background in many X-ray astronomy missions \cite{hall2007modelling}, in many cases even forming the dominant component of background. The background count rate induced by such particles in a particular mission can be expected to depend on many factors, including the detector design and the spacecraft structure surrounding a detector. In general, the X-ray-like background continuum in thinner detectors would be expected to be composed of a higher proportion of secondary electron-induced background than thicker detectors, primarily because thicker detectors are more susceptible to Compton scattering photons. However, simulation results presented in Chapter 8 of Davis (2021)\cite{davis2021simulation} indicate that background induced by secondary electrons either through absorption or backscattering still accounts for approximately two thirds of the total background even in a simulated detector with a 450 micron thick depletion region. This means that removing the secondary electron component of X-ray-like background could remove the majority of X-ray-like background in many space-based X-ray astronomy missions, particularly in spectral regions where on-axis background is not significant, and in missions where mitigation strategies are being employed to remove on-axis background. Reducing this background would improve signal-to-noise ratios of scientific observations, reduce the time needed to make scientific observations, and reduce systematic errors for any observations made by a detector.

Research has also been performed to investigate potential mechanisms for magnetic diversion of incoming cosmic rays themselves for both space astronomy missions in general as well as for human-manned missions\cite{westover2019magnet}, some of which are somewhat exotic designs such as the use of confined plasmas\cite{bamford2014exploration}. Designing a system that can do this is a significantly more challenging task than deflecting electrons and soft protons, as cosmic rays capable of inducing background and causing radiation damage have significantly larger momenta/magnetic rigidities and approach the detector from all directions. This means that such a system would require a very strong magnetic field encompassing the spacecraft, which would likely need to be generated by strong superconducting magnets.

However, only minimal research has been done to investigate the use of magnetic shielding to specifically shield X-ray detectors from secondary particles arriving omnidirectionally, which is a very different problem to the problem of shielding a detector from particles arriving from approximately a single direction. Some research investigating the ability of magnetic shielding to repel secondary electrons in an X-ray detector has however previously been performed by Perinati et al. (2018) \cite{perinati2018magnetic}. 

Perinati et al. introduce some initial concepts for a potential omnidirectional secondary electron magnetic shielding design using the theoretical case of a ring-shaped permanent magnet embedded into the walls surrounding a detector (in the example case of the ATHENA WFI, this is an aluminium proton shield). Perinati et al. then used the approximated magnetic field for the configuration in magnetic field trajectory simulations to calculate several electron trajectories for kinetic energies between 1.7~keV and 14.8~keV originating at several points on a surface near the detector, showing that the magnetic field used is in principle capable of deflecting background-inducing electrons. Perinati et al. also investigated the magnetic field that such a configuration would generate across a detector, due to concerns that a magnetic field might adversely affect detector functioning, finding that the magnetic field would be on the order of several milliTesla. However, the experiments described in Chapters 6 and 7 of Davis (2021)\cite{davis2021simulation}, which are also discussed in Hall et al. (2018)\cite{hall2018predicting} and Eraerds et al. (2021)\cite{eraerds2021enhanced}, have since indicated that a magnetic field of several milliTesla incident upon a CCD97 X-ray detector do not appear to cause significant changes to detector functionality.

The first half of Chapter~10 of Davis (2021)\cite{davis2021simulation} expands on the ideas presented by Perinati et al., describing a lot of the physics underpinning active magnetic shielding for the diversion of secondary background-inducing electrons and the strategies for designing these types of magnetic shielding. A mathematical framework is also developed for calculating the effectiveness of uniform magnetic fields of reducing secondary electron background, which some readers may find useful although not strictly necessary, for understanding the results presented in this article.

\section{Simulation Design}

To advance the technology readiness of the hypothetical concept of omnidirectional secondary electron magnetic shielding, the simulations described in this article were designed to test the full functionality of a magnetic shielding configuration on an actual representation of the radiation environment of space. This meant utilising a full Galactic Cosmic Ray (GCR) proton spectrum that would be incident upon an approximate detector structure isotropically, to assess the full effect of magnetic shielding on the X-ray-like background continuum.

\begin{figure}
    \centering
    \includegraphics[]{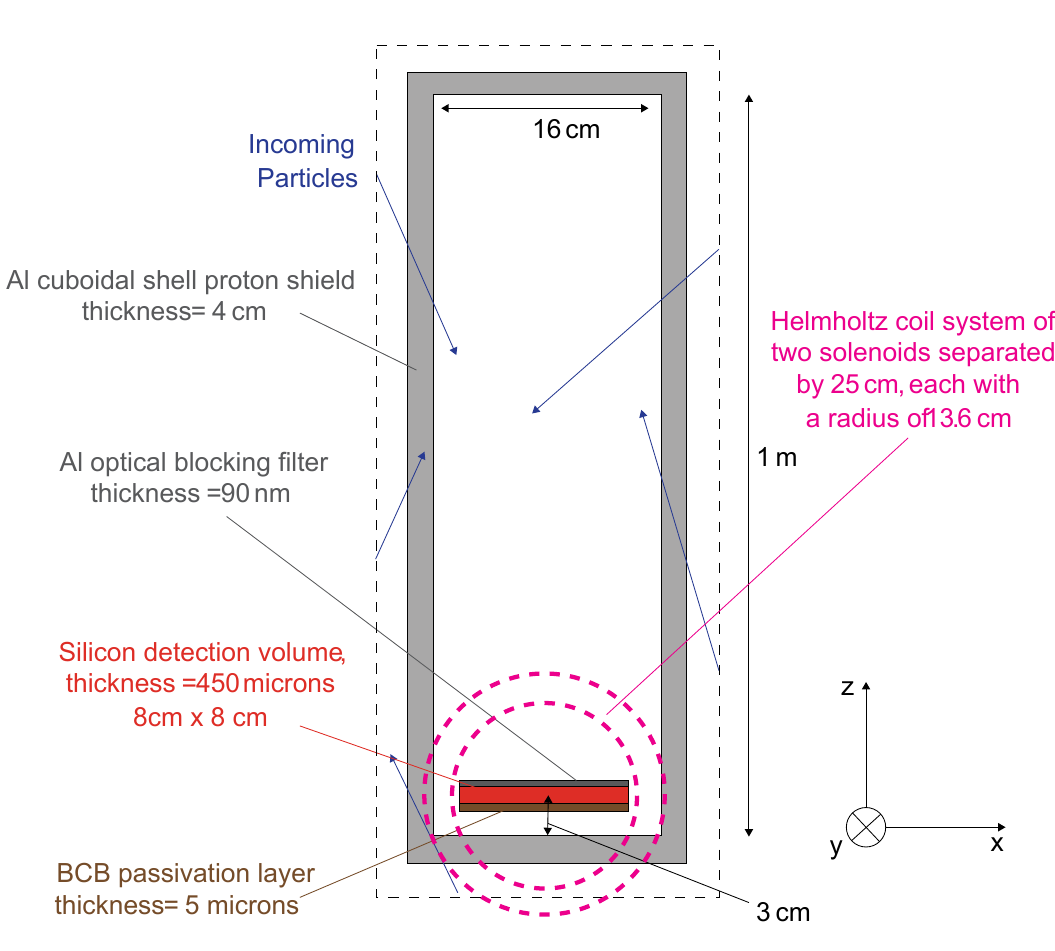}
    \caption{The geometry used for the simulations described in this article. The geometry used was cuboidal, in contrast to the spherical shell geometries that are frequently used, so that the system might more accurately reflect a real geometry that might surround a spacecraft detector. This was particularly important in these simulations because the applied magnetic field would break many of the symmetries that are often present in spherical shell spacecraft simulation models. The main volume of importance to the generation of secondary particles is the `proton shield' surrounding the detector region, which was set to be 4~cm of aluminium in thickness.}
    \label{fig:simGeometry}
\end{figure}

To assess the effectiveness of a magnetic field at shielding a detector from a particular electron energy, Geant4\cite{agostinelli2003geant4} simulations were needed to fully quantitatively assess the effectiveness of magnetic shielding at reducing X-ray-like background. Two sets of simulations were performed to examine the effect of a Helmholtz coil magnetic field on X-ray-like background, each of which used the cuboidal geometry and Helmholtz coil magnetic field configuration as displayed in \cref{fig:simGeometry}. The overall detector structure was designed to be analogous to current designs for the Wide Field Imager (WFI) on ESA's ATHENA mission \cite{von2018evaluation,eraerds2021enhanced} at the time of writing, which is currently expected to have a 450~micron thick depletion region, and a 90~nm thick on-chip optical blocking filter.

Here the same magnetic field was utilised as in the experiments and simulations described in Chapters 6 and 7 of Davis (2021) \cite{davis2021simulation}, with a simulated DC current that could be set to be between 0A and 60A passing through the ‘wires’. The Helmholtz coils these were designed to simulate were Teltron Helmholtz Coil D solenoids\cite{TeltronCoils2015} albeit with a radius of 13.6~cm as opposed to 6.8~cm, where current in this article describes the current through each wire loop in a solenoid rather than the bulk current through the whole solenoid at once. The simulated Helmholtz coils were placed such that they were separated by 25~cm on each side of the detector. It should be noted that the magnetic field strength could be increased significantly around the detector in this case if the coils were moved closer together inside the proton shield here.

In the experiments described in Chapters 6 and 7 of Davis (2021)\cite{davis2021simulation}, the maximum current through the solenoids used was 2A; however, this was only used as the maximum to prevent the epoxy insulation surrounding coil wires from becoming damaged due to high temperatures created by Ohmic heating. For solenoids to be used on a spacecraft, it would not be expensive to use wires that could withstand significantly higher temperatures and currents than the coils used in those experiments, as the coils that were used in the experiments were inexpensive solenoids that were initially designed for educational purposes. 

Cooling mechanisms could also be implemented to keep the wires from overheating, which would also have the effect of lowering the resistance in the wires, reducing the amount of power that would be required to keep the coils operating at a particular current. Cooling mechanisms are commonly implemented for many applications in spacecraft, for example, the ATHENA XIFU is expected to include a Detector Cooling System (DCS) that is capable of reducing temperatures to as low as several tens of milliKelvin \cite{prouve2018athena}. 

The power requirements for a Helmholtz coil of the design described in this article are given in \cref{tab:powerReqTable} for room temperature, where the wire area was set to be approximately 0.0024 mm$^2$, the same wire area as the wires in Teltron Helmholtz Coil D solenoids \cite{TeltronCoils2015}. In addition to the power requirements for copper wire solenoids, the power requirements for aluminium wire solenoids are also given. Aluminium has a lower atomic number than copper, making it viable as an alternative to copper with fewer fluorescence lines and lower density, at the expense of having higher resistivity.

\begin{table}
\caption{\label{tab:powerReqTable} The magnetic field strength at the central point of the simulated Helmholtz coil system for each current and the required power at room temperature to achieve certain currents in the simulated Helmholtz coil system that was used. While the power requirements in these cases are reasonable in size, it is anticipated that these power requirements could be significantly reduced through many potential solutions, including modifying the coil structure, reducing the temperature of the coils - perhaps using cooling mechanisms, modifying the geometry around the detector, using more conductive or even superconductive materials in the coils. Alternatively replacing the coils with permanent bar magnets would remove the need for power altogether.}
\centering
\begin{tabular}{|p{0.2\textwidth}|p{0.25\textwidth}|p{0.25\textwidth}|p{0.25\textwidth}|}
    \hline
    \begin{center}Current (A)\end{center} & Magnetic field strength at central point of solenoids & Power requirement at room temperature for copper wire solenoids & Power requirement at room temperature for aluminium wire solenoids \\
    \hline
    \centering 2 & 4.7 mT & 0.192 kW & 0.294 kW \\
    \hline
    \centering 4 & 9.4 mT & 0.768 kW & 1.175 kW \\
    \hline
    \centering 6 & 14.2 mT & 1.728 kW & 2.643 kW \\
    \hline
    \centering 8 & 18.9 mT & 3.072 kW & 4.698 kW \\
    \hline
    \centering 10 & 23.6 mT & 4.800 kW & 7.341 kW \\
    \hline
\end{tabular}
\hspace{0mm}
\end{table}

The exact specifications of any magnetic system would have to be chosen in conjunction with spacecraft and electronic engineering considerations. The presence of the produced magnetic field alone is not expected to have any negative effect on the scientific quality of observations as soft protons entering the system through X-ray optics would be somewhat deflected by a magnetic shielding system just like secondary electrons. Additionally, simulations given later in this article indicate that further secondaries produced by secondary particles hitting spacecraft walls due to magnetic field deflection are not significant for inducing background. However, as with any equipment added to a spacecraft design a magnetic diversion system will somewhat increase spacecraft mass and complexity, and introduce volume constraints around surrounding devices. Therefore the inclusion and design of any such system would need to assessed versus engineering constraints and considerations, and the advantages of improved science quality may have to be weighed up against engineering challenges introduced by different magnetic shielding designs.

Care would also have to be taken to ensure that devices within the magnetic field are not sufficiently affected by the magnetic field. As discussed earlier, the presence of Teltron Helmholtz coil magnetic fields surrounding a CCD97 in experiments did not affect device functionality, however it is possible in principle that some electronics may be affected by magnetic fields (such as certain storage devices for instance). In some missions this may introduce further constraints and engineering challenges, although the presence or scale of these potential challenges depends heavily on mission spacecraft design.

While the power requirements given in \cref{tab:powerReqTable} are relatively large, it is foreseen that through cooling mechanisms, and through optimisation by increasing the solenoid wire area or decreasing solenoid radius, the power requirements given here could be significantly reduced. The required current could also be reduced through a reduction in separation between solenoids, which would significantly increase the central magnetic field strength for a given current. This design could also be replicated using permanent magnets, which would remove the power requirements for the system entirely.

Simulations were set up using the parameters described in \cref{tab:Geant4settings}. The use of these parameters have been found to produce accurate Geant4 simulations of background-inducing particles at keV energies\cite{davis2021simulation}. The electromagnetic stepper used was the default stepper for Geant4 version 10.4.p01.

\begin{table}
\caption{\label{tab:Geant4settings} Some of the settings used in the Geant4 simulations described in this article.}
\centering
\begin{tabular}{|p{0.4\textwidth}|p{0.4\textwidth}|}
    \hline
    \begin{center}Setting\end{center} & \begin{center}Value\end{center}\\
    \hline
    Geant4 version & 10.4.p01 \\
    \hline
    Physics List & QGSP\_BIC\_HP\_PEN \\
    \hline
    Default production cut length & 1 \textmu m \\
    \hline
    Minimum energy & 0.1 keV \\
    \hline
    PIXE flag & True \\
    \hline
    Auger flag & True \\
    \hline
    Fluo flag & True \\
    \hline
    Scattering type & Multiple scattering \\
    \hline
    fMinStep & 0.01 mm \\
    \hline
    fDeltaChord & 0.002 mm \\
    \hline
    fDeltaOneStep & 0.01 mm \\
    \hline
    fDeltaInterSection & 0.1 mm \\
    \hline
    fEpsMin & 0.25 nm \\
    \hline
    fEpsMax & 0.005 mm \\
    \hline
\end{tabular}
\hspace{0mm}
\end{table}

The first set of simulations that were performed were designed to examine representative general trajectories that electrons would take within the volume surrounding the detector when the magnetic field was switched on. As these were only meant for examining the effect of the magnetic field on particle trajectories from a qualitative perspective, all volumes were set to be vacuum, and primary particles were set to be 7~keV electrons, with initial momentum directions set to be isotropically distributed. Simulations were performed using both no magnetic field and the magnetic field corresponding to a current of 2A passing through the Helmholtz coils. The results for these simulations are presented in \cref{sec:represent7keVElecSims}.

The second set of simulations were designed to be full simulations of how the magnetic field of the Helmholtz coil system would influence the X-ray-like background in the real environment of space. Therefore, all the volume materials for these simulations were switched back from vacuum to the actual solid volume materials as given in \cref{fig:simGeometry}. As knock-on electrons that induce background are primarily produced by cosmic protons, the simulations were run using the Galactic Cosmic Ray (GCR) proton input spectrum as described by the CREME96 model\cite{tylka1997creme96}. The full event detection and normalisation process that was used for these simulations is the same as that described in Chapters 5 and 8 in Davis~(2021)\cite{davis2021simulation}. Simulations were run for many Helmholtz coil currents between 0A and 60A, and valid X-ray-like background hits were defined to be any particle hits that caused energy depositions between 0.1~keV and 15~keV, while not creating a particle track. The results for these simulations are presented in \cref{sec:FullCosmicProtonSims}.

\section{Representative 7~keV Electron Simulation Results}
\label{sec:represent7keVElecSims}
\begin{figure}
    \centering
    \newcommand{\fourFigHeight}{12cm}
    \subfloat[The geometry only]{
        \label{fig:noParticles}\includegraphics[height=\fourFigHeight]{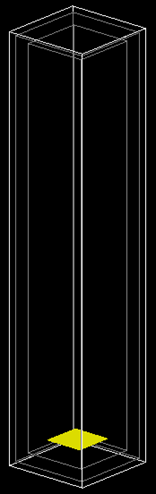}
    }
    \subfloat[no magnetic field]{
        \label{fig:noMagField}\includegraphics[height=\fourFigHeight]{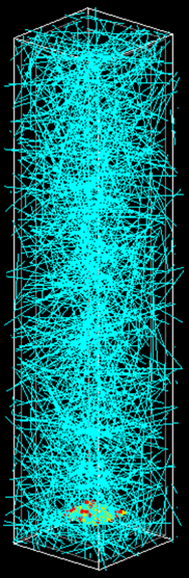}
    }
    \subfloat[with magnetic field]{
        \label{fig:withMagField}\includegraphics[height=\fourFigHeight]{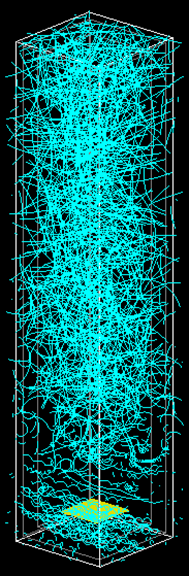}
    }
    \subfloat[with magnetic field, side-on viewing angle]{
        \label{fig:withMagFieldSideView}\includegraphics[height=\fourFigHeight]{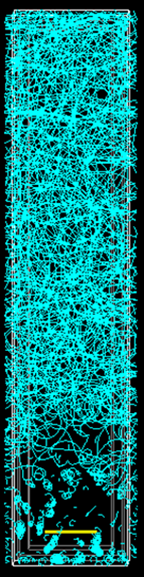}
    }
    \caption{The Geant4 geometry used in all the simulations discussed in this paper, as described in \cref{fig:simGeometry}. \Cref{fig:noParticles} displays the geometry used in the simulations, with no particle tracks displayed. \Cref{fig:noMagField} displays the particle tracks from a simulation of 5~keV electrons incident upon the geometry, where all materials in the geometry have been temporarily set to be vacuum. Red circles indicate points where particle trajectories intersected with the volume corresponding to the `detector' location. Finally, \cref{fig:withMagField,fig:withMagFieldSideView} are two different viewpoints of another simulation which used exactly the same simulation set-up as shown in \cref{fig:noMagField}, but where the magnetic field of a Helmholtz coil with a current of 2A has been applied across the geometry, as described in \cref{fig:simGeometry}. It can be seen in \cref{fig:withMagField,fig:withMagFieldSideView} that the presence of the magnetic field creates a region around the detector where particles are much less likely to enter.}
    \label{fig:geomTrajComparisons}
\end{figure}

The Geant4 trajectories of the 7~keV electrons for the first set of simulations in the Helmholtz magnetic field structure are displayed in \cref{fig:geomTrajComparisons}. 

It can be seen in \cref{fig:geomTrajComparisons} that a region with a significantly lower density of electrons is generated around the detector when the magnetic field is switched on, indicating that a Helmholtz coil magnetic field is indeed effective at shielding the detector from particles. This region of lower electron density is the region directly between the Helmholtz coils, and therefore the region of the simulation with the strongest magnetic field strength and electron trajectory curvatures.

Electrons entering this region appear to be undergoing one of two behaviours. One possibility is that the electron enters the region but due to the increased curvature of its trajectory within the stronger magnetic field strength the electron's momentum gets rotated around and the electron becomes quickly deflected away from the region. The other possibility is that the electron becomes confined to one of the field lines, in which case the particle follows one of the field lines while rotating around it. In either situation, it is challenging for a particle to reach the detector and create an X-ray-like background hit, which explains why significantly less particles are registered as impacting the detector when the magnetic field is switched on in \cref{fig:withMagField} as opposed to when the magnetic field is switched off in \cref{fig:noMagField}.

\section{Full Cosmic Ray Proton Spectrum Simulation Results}
\label{sec:FullCosmicProtonSims}

\begin{figure}
    \centering
    \includegraphics[]{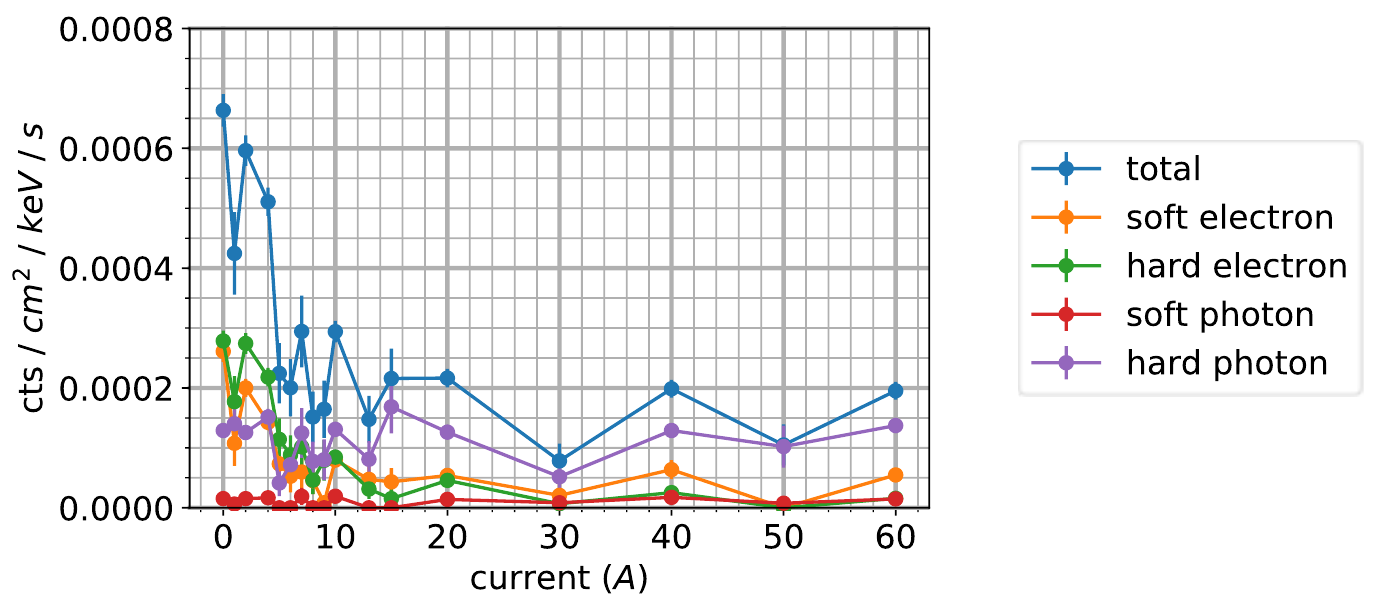}
    \caption{The off-axis X-ray-like background for deposited energies between 2~keV and 7~keV as a function of the simulated Helmholtz coil current. The increase in Helmholtz coil current significantly reduces the X-ray-like background in the detector. This reduction in X-ray-like background comes from the reduction in both the soft and hard/backscattering electron components of background. Here, 'soft' particles are defined as particles that deposit their full energy into the detector, and 'hard' particles are particles that only deposit some of their incident energy. Hard electrons are usually backscattering electrons, and hard photons are usually Compton scattering photons\cite{davis2021simulation}.}
    \label{fig:fluxVsCurrent}
\end{figure}

The results for the full simulations using GCR protons as the input spectrum, and with volumes in the geometry set to solid materials are displayed in \cref{fig:fluxVsCurrent}, where the mean count rate between 2~keV and 7~keV is plotted as a function of the current through the coils. The mean count rate appears to significantly reduce as the current is increased, showing that the Helmholtz coil configuration used is successful at reducing background significantly, and is effective as a magnetic shielding mechanism.

The reduction in count rate is significant for both soft electrons and hard electrons, even though the median kinetic energy for background-inducing hard electrons is significantly higher than the median kinetic energy for soft electrons\cite{davis2021simulation}. 

This may be explained by the fact that when the magnetic field is perpendicular to the surface than an electron is generated from (as is the case for many of the walls in the geometry given here), the median distance a particle will travel from its generation surface in a magnetic field is reasonably small relative to the maximum possible distance it can possibly travel\cite{davis2021simulation}. An alternate explanation may also be that the angle at which backscattering electrons are most likely to deposit within the detector energy range is such that a relatively weak magnetic field is capable of preventing electrons from generating an X-ray-like event. This explanation might be evidenced by some of the experimental and simulation results given in Chapter 7 of Davis (2021)\cite{davis2021simulation}. 

In Chapter 7 in Davis (2021)\cite{davis2021simulation} it was found that a wide region of produced spectra were composed of high energy, fully-penetrating electrons. This region of the spectrum did not vary significantly with changes in magnetic field strength during the 200~MeV experiment or in the simulations for that experiment, and was primarily composed of particles that were in the region of approximately 100~keV in kinetic energy, which is similar to common energies of backscattering electrons as found in Chapter 8 of Davis (2021) \cite{davis2021simulation}. A major difference between the geometry described in Chapter 7 in Davis (2021)\cite{davis2021simulation} and the simulations in this article is that in the former case particles were almost all incident upon the detector approximately vertically when no magnetic field was present. In that instance, the magnetic field was not strong enough to divert fully-penetrating particles away from the detector.

However the results given in \cref{fig:fluxVsCurrent} indicate that in a simulation much more similar to the actual space-based environment, the Helmholtz coils are very effective at removing X-ray-like background due to backscattering electrons. This indicates that the removal of background here may be in some way related to the specifics of the backscattering angular cross-section, and angle of incidence for backscattering electrons impacting the detector.

In addition to investigating the mean count rate across 2 keV – 7 keV as a function of current, it is instructive to determine how the spectrum changes with Helmholtz coil current. The spectrum for several currents is plotted in \cref{fig:spectraForCurrents} below, where it appears that the presence of the magnetic field appears to flatly reduce the continuum spectrum without significantly altering the shape of the spectrum.

\begin{figure}
    \centering
    \includegraphics[]{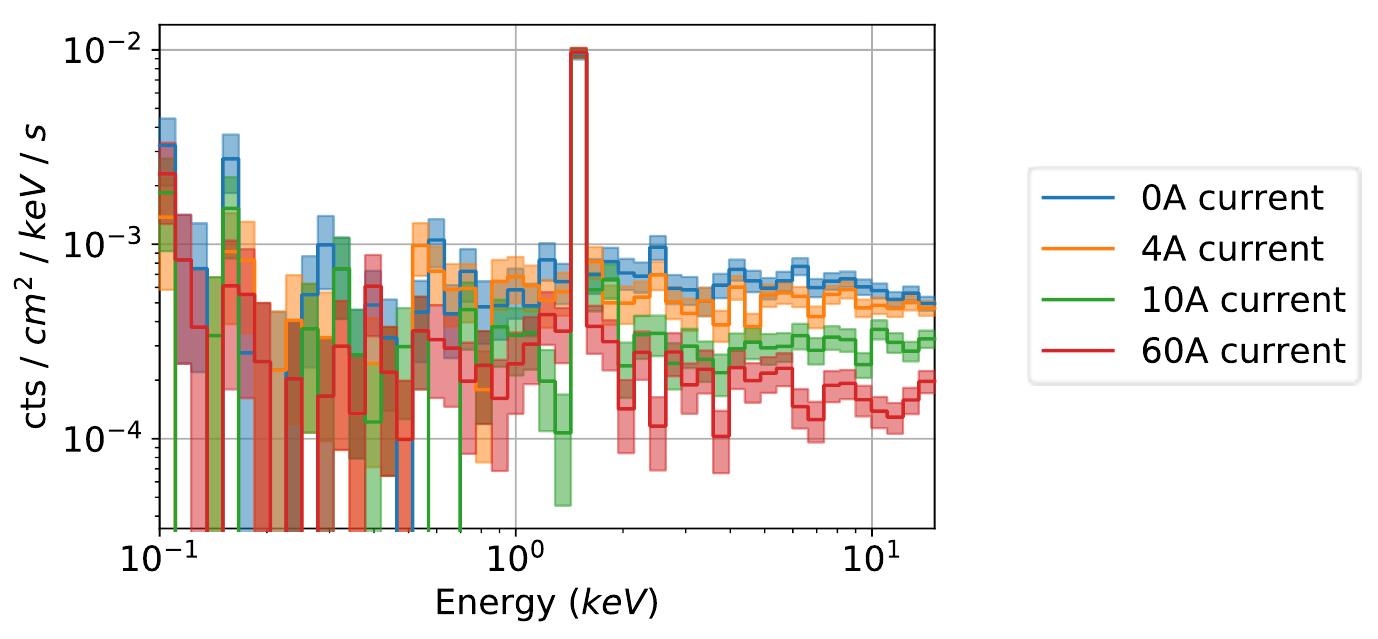}
    \caption{The spectrum associated with the simulated X-ray-like background for several of simulated Helmholtz coil `currents'. The background continuum reduces across the whole energy range (which is the approximately the expected energy range of the ATHENA WFI). The large increase in count rate for the bin corresponding to 1.48~keV is the fluorescence line of aluminium.}
    \label{fig:spectraForCurrents}
\end{figure}

\begin{figure}
    \centering
    \includegraphics[]{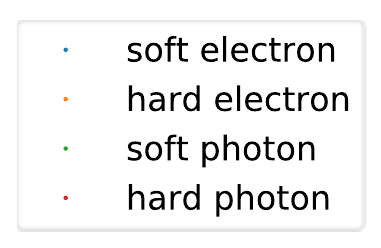}
    \subfloat[No magnetic field]{
        \includegraphics[width=0.45\linewidth]{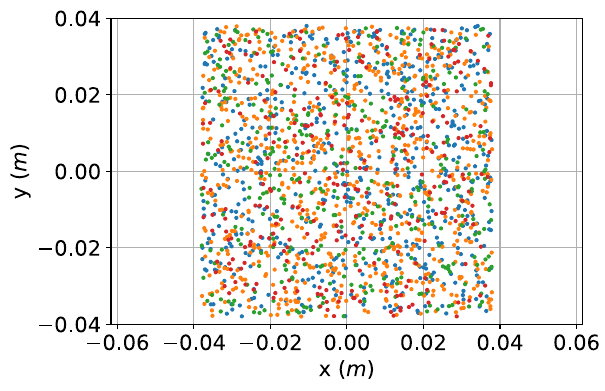}
    }
    \subfloat[With magnetic field]{
        \includegraphics[width=0.45\linewidth]{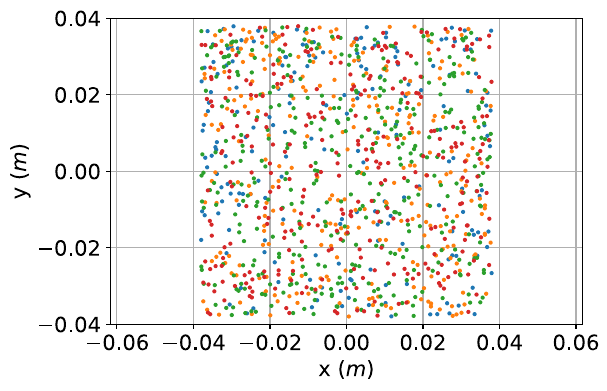}
    }
    
    \caption{The distribution of absorbed X-ray-like background hits across the simulated detector. The presence of the magnetic field does not seem to create any significant patterns of background across the simulated detector. Here, 'soft' particles are defined as particles that deposit their full energy into the detector, and 'hard' particles are particles that only deposit some of their incident energy. Hard electrons are usually backscattering electrons, and hard photons are usually Compton scattering photons, as discussed in Davis (2021)\cite{davis2021simulation}.}
    \label{fig:spatialDistBkg}
\end{figure}

The spatial distribution of X-ray-like events across the detector for the 0 A and 10 A cases are also displayed in \cref{fig:spatialDistBkg}, where the distribution appears to be approximately spatially uniform. This means that the introduction of a Helmholtz magnetic field would not appear to introduce systematic errors associated with background sampling across images.

\begin{figure}
    \centering
    \includegraphics[]{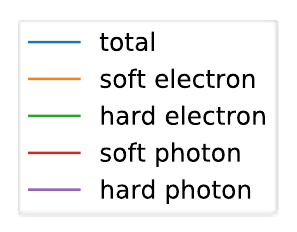}
    \newcommand{\gridFigWidth}{0.45\linewidth}
    \subfloat[No magnetic field, linear scale]{
    \label{fig:bkgZaxisA}
    \includegraphics[width=\gridFigWidth]{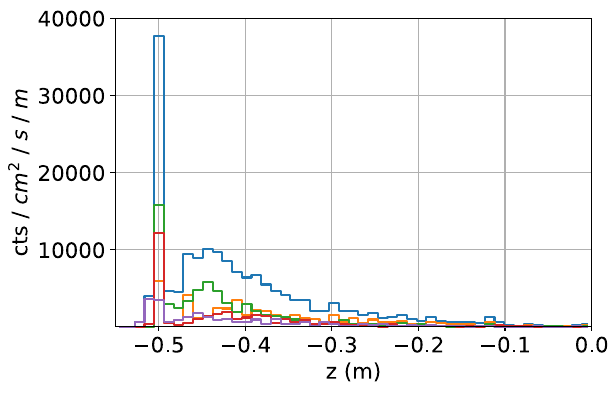}
    }
    \subfloat[With magnetic field, linear scale]{
    \label{fig:bkgZaxisB}
    \includegraphics[width=\gridFigWidth]{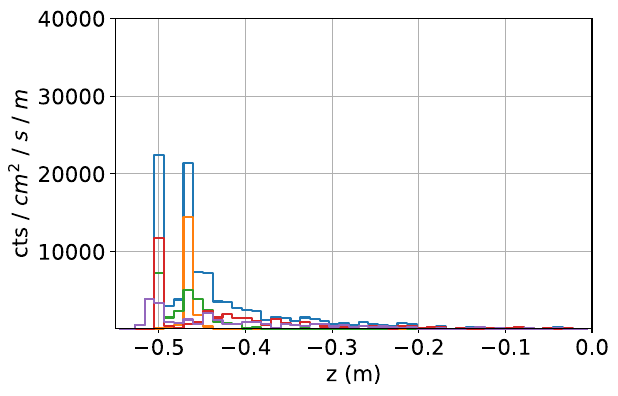}
    }
    \hspace{0mm}
    \subfloat[No magnetic field, logarithmic scale]{
    \label{fig:bkgZaxisC}
    \includegraphics[width=\gridFigWidth]{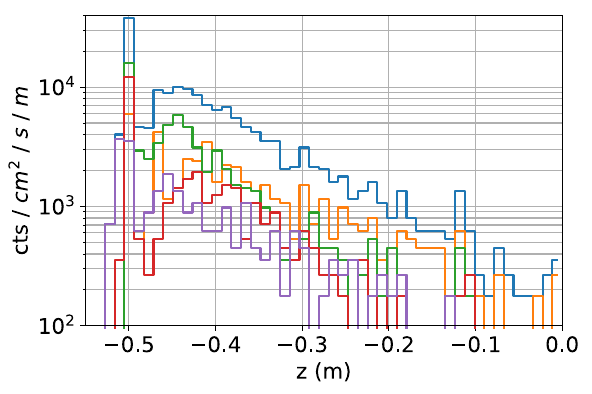}
    }
    \subfloat[With magnetic field, logarithmic scale]{
    \label{fig:bkgZaxisD}
    \includegraphics[width=\gridFigWidth]{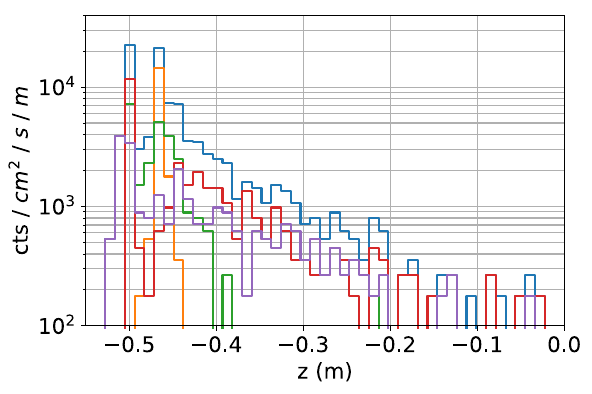}
    }
    \hspace{0mm}
    \caption{The differential flux of particles that generate X-ray-like background hits in the simulated detector for deposited energies between 2~keV and 7~keV, as a function of position along the z-axis. As expected, the presence of the magnetic field significantly reduces the quantity of particles reaching the detector from surfaces both close to and a moderate distance away from the detector. As the quantity of particles originating from more distant locations is reduced more than that of particles closer to the detector, positioning the detector further away from other surfaces within the spacecraft geometry while the magnetic field is present will reduce the X-ray-like background even further. Alternatively this may allow for a weaker magnetic field to be used with lower power requirements. Here, 'soft' particles are defined as particles that deposit their full energy into the detector, and 'hard' particles are particles that only deposit some of their incident energy. Hard electrons are usually backscattering electrons, and hard photons are usually Compton scattering photons\cite{davis2021simulation}.}
    \label{fig:bkgZaxis}
\end{figure}

It is also instructive to investigate the spatial distribution of particle vertex positions, particularly the distribution of X-ray-like background inducing origin locations along the baffle axis, which is displayed in \cref{fig:bkgZaxis}.

\Cref{fig:bkgZaxis} indicates that as expected, the magnetic field decreases the number of electrons originating from more distant surfaces to the detector more effectively than from surfaces closer to the detector. In particular, it can be seen that a significant proportion of background counts in \cref{fig:bkgZaxisB} and \cref{fig:bkgZaxisD} originate from the surface behind the detector, which is only 3 cm away from the detector itself. Some of these are hard electron counts, meaning the effectiveness of the Helmholtz magnetic shielding may be further improved by suspending the detector a larger distance away from the back surface of the baffle, or by adding more material underneath the detector to block knock-on electrons.

It can also be seen that a significant number of soft electron background counts in \cref{fig:bkgZaxisB} and \cref{fig:bkgZaxisD} originate from the detector volume itself at -0.47 m, which are not present in the plots for no magnetic field. This implies that these counts are produced from the detector itself and are spiralling backwards in the magnetic field back into the detector. As these particles will likely hit the detector in locations near to the pixel from which they were produced by some high energy particle, it is likely that self-anticoincidence methods\cite{burrows2019us,bulbul2020characterization} could be used to remove these electrons from background images. This would increase the effectiveness of the magnetic shielding even further.

Overall the Helmholtz magnetic field design appears to be very effective at removing electron-induced X-ray-like background from images. The effectiveness of this magnetic shielding could be optimised from this simple design in several ways, including by for instance increasing the distance between the detector and surfaces within line of sight of the detector faces and increasing the thickness of on-chip layers. It is also possible that more complex magnetic fields may produce further increases in effectiveness. These might be achieved by the addition of more solenoids to the geometry.

It may also be possible to replicate the field of the solenoids here using permanent magnets, perhaps by using single large magnets for each solenoid that have a surface current such that they replicate the effect of a coil of wires. Alternatively, it may be possible to replicate a solenoid field using a ring of magnets for each solenoid similar to that used by primary particle magnetic diverters in existing missions \cite{ferreira2018design}. While passive magnetic shielding like this would remove the power requirements of generating such a magnetic field, the magnets may need to be graded-Z shielded \cite{atwell2013mitigating} to ensure that fluorescence from high atomic number materials in the magnets does not influence the detector background.

Additionally, combining magnetic shielding with a graded-Z shielding configuration that is capable of attenuating hard photons produced by GCR protons and self-anticoincidence algorithms means that it may be possible to significantly reduce the GCR proton-induced background, as well as background induced by the CXB, GCR electron and GCR alpha-induced background. However, further research into hard photon generation and graded-Z shielding is necessary before this might be achieved.

\section{Conclusions}

Geant4 simulations performed using a Helmholtz coil magnetic field with a cuboidal proton shield structure show that magnetic shielding may be very effective at reducing the background induced by omnidirectional low energy or `soft' and high energy or `hard' electrons. This indicates that a Helmholtz coil magnetic field is likely to be an effective active magnetic shielding design for shielding a detector from all knock-on electrons. A magnetic field of this structure may be achieved by either permanent magnets or electromagnets providing that the power requirements for an electromagnetic configuration can be kept sufficiently low, perhaps through cooling mechanisms, or that any fluorescence from high atomic number permanent magnetic configurations can be minimised.

\subsection*{Disclosures}

The authors declare that they have no relevant or material financial interests that relate to the research described in this paper. 

\subsection* {Acknowledgments}
The first author would like to thank everyone at the Center for Electronic Imaging at The Open University, the institution they were studying at during their PhD, during which time this research was performed. They would particularly like to thank their supervisor at the time, Dr. David Hall, for all his advice and help, as well as Dr. Michael Hubbard, Dr. Neil Holyhoke and Oliver Hetherington for discussions about this research. This research was funded by STFC and Teledyne-e2v.



\bibliography{report}   
\bibliographystyle{spiejour}   



\vspace{1ex}


\end{spacing}
\end{document}